\documentclass[12pt]{article}
\usepackage{latexsym}
\usepackage{epsfig,graphics}

\newcommand{\be}{\begin{equation}}
\newcommand{\ee}{\end{equation}}

\def\lsi{\raise0.3ex\hbox{$<$\kern-0.75em\raise-1.1ex\hbox{$\sim$}}}
\def\gsi{\raise0.3ex\hbox{$>$\kern-0.75em\raise-1.1ex\hbox{$\sim$}}}

\begin{document}

\begin{titlepage}

\begin{centering} 

\null\vspace{2.5cm}

{\bf 
VORTICES AND CONFINEMENT IN\\
HOT AND COLD D=2+1 GAUGE THEORIES}

\vspace{1.8cm}

A. Hart$^{\rm a}$, B. Lucini$^{\rm b}$, 
Z. Schram$^{\rm c}$ and M. Teper$^{\rm b}$

\vspace{0.6cm}
{\em $^{\rm a}$%
Department of Physics and Astronomy, University of Edinburgh,\\
Edinburgh EH9 3JZ, Scotland, UK\\}
\vspace{0.3cm}
{\em $^{\rm b}$%
Theoretical Physics, University of Oxford, 1 Keble Road, \\
Oxford OX1 3NP, England, UK\\}
\vspace{0.3cm}
{\em $^{\rm c}$%
Department of Theoretical Physics, University of Debrecen, \\
H-4010 Debrecen P.O.Box 5, Hungary\\}

\vspace{1.5cm}
{\bf Abstract.}
\end{centering}

\noindent
We calculate the variation with temperature of the vortex 
free energy in D=2+1 SU(2) lattice gauge theories. We do so 
both above and below the deconfining transition at $T=T_c$. 
We find that this quantity is zero at all $T$ for large enough
volumes. For $T < T_c$ this observation is consistent with 
the fact that the phase is linearly confining; 
while for $T > T_c$ it is consistent with the conventional 
expectation of `spatial' linear confinement. In small 
spatial volumes this quantity is shown to be  non-zero.
The way it decreases to zero with increasing
volume is shown to be controlled by the (spatial) string
tension and it has the functional form one would 
expect if the vortices being studied were responsible for
the confinement at low $T$, and for the `spatial' confinement
at large $T$. We also discuss in detail some of the direct numerical
evidence for a non-zero spatial string tension at high $T$,
and we show that the observed linearity of the (spatial) 
potential extends over distances that are large compared to
typical high-$T$ length scales.

\vfill

\end{titlepage}

\section{Introduction}
\label{sec_intro}

The idea that vortices are the relevant degrees of 
freedom for understanding confinement in non-Abelian
gauge theories is an old one 
\cite{thooft}.
Consider, for example, an arbitrary SU(2) gauge field,
$\{A_\mu(x)\}$,
in three dimensional Euclidean space-time. Now suppose
that we perform the following gauge transformation
$\{A\} \to \{A^g\}$.
Choose a space-like plane and choose a point $x_c$
in that plane. Consider any closed loop that encloses 
that point. We perform a
gauge transformation such that, as we go around
that closed loop, the gauge transformation varies
from $1$ to a non-trivial element of the centre of the
group: in this case $z=-1$. Clearly this gauge transformation
is not single-valued; nonetheless the gauge potential is, because
it is in the adjoint representation. However any field
in the fundamental representation would acquire a factor 
of $-1$; in particular, the Wilson loop defined on such 
a closed curve, $c$, will acquire a factor of $-1$ relative
to its value prior to the gauge transformation
\be
W_c[A^g] = Tr \ e^{\int_c A^g \cdot ds} = - W_c[A]    
\ee
We extend this construction to all space-like planes,
so that the point $x_c$ traces a line 
and we shall refer to this
as the vortex (world) line. We can easily generalise the
construction so that this line may be closed rather than 
infinite. If the closed curve along which the Wilson loop is 
defined links once with the vortex line, then the Wilson loop 
will acquire a factor of $-1$. 
If it links $N$ times then it acquires a factor
of $(-1)^N$. One can consider the field $\{A^g\}$
to be the original field  $\{A\}$ with an added ultraviolet
centre vortex, whose core coincides with the vortex line.
Along this line the gauge transformation is singular and the
gauge potentials are ill-defined. Since such a vortex
corresponds to a gauge transformation everywhere
except along a line, we can readily define fields
with any number of such vortices (since the problematic
vortex intersections will have measure zero). 
And it is then straightforward to show, as we later shall,
that if the vacuum contains an uncorrelated 
gas of arbitrarily long vortices then the average value 
of a Wilson loop bounded by a planar curve $c$, will decay 
as the exponential of its (minimal) area, ${\cal A}[c]$,
\be
\langle W_c \rangle \propto    \exp\{-\sigma {\cal A}[c]\}
\label{eqn_wl_area}
\ee
with the resulting string tension, $\sigma$  being proportional
to the average number of vortex lines threading 
each unit of area.

When the theory is regularised, such an ultraviolet
vortex will become well-defined but will have a 
divergent action density along the actual vortex line. 
For example, consider a cubic lattice 
with a plaquette action. If we place the vortex line 
so that it avoids all sites and links, then the field
is clearly well-defined everywhere. However the 
plaquettes threaded by this line will have their 
signs flipped, and this will correspond to an
action density along the vortex line, 
that diverges as we approach
the continuum limit. Quantum fluctuations and
renormalisation will presumably lead to the line
spreading into a tube of finite width, with a non-zero
but finite action density within this vortex `core'.
Indeed, at the classical level
we would expect this core to spread to
fill the whole box: in which case such a vortex ceases 
to be a useful degree of freedom. However it may be
that quantum fluctuations induce an effective potential 
for these vortices such that the core is stabilised with 
a width given by some characteristic length scale.
Moreover if this potential is such that these vortices 
also condense into the vacuum, then they will produce
linear confinement between particles in the 
fundamental representation. But all this is, of course,
not guaranteed; if, for example, the vortices are 
themselves confined, then they will not disorder
large Wilson loops. 

Whether such vortices exist and, if so, whether they condense, 
are thus the important dynamical questions for this
approach to confinement. They are also difficult non-perturbative 
questions that have begun to be addressed, during
the last two years, in a number of lattice Monte Carlo studies.
(See, for example,
\cite{greensite99,kovacs00}
and references therein.) 

Linear confinement corresponds to an area decay, as in 
(\ref{eqn_wl_area}), for arbitrarily large Wilson loops.
If a vortex is to disorder such a loop it must pierce the 
minimal area an odd number of times and this implies that 
the vortex line must be correspondingly long. (Unless it
is close to the perimeter.) That is to say, one needs a 
condensate of arbitrarily long vortex lines. Actually this
is only so at $T=0$. 
At high $T$ the requirement becomes far less severe. A vortex 
that closes upon itself through the periodic temporal 
direction, is only of length $O(1/T)$ and yet can disorder
spatial Wilson loops that are arbitrarily large. (It will
not disorder Wilson loops that lie in a space-time plane,
but that is to be expected in the high-$T$ deconfining phase.) 
If vortices exist at all then there will surely exist a finite
density of such periodic vortices in the high-$T$ (Euclidean) 
vacuum. That such short periodic vortices might be the dynamics 
underlying the apparent
area decay of spatial Wilson loops at high $T$ is an old idea. 
However if the interactions between these vortices were
such that they totally screened each other, or were confined, 
then they would not disorder large spatial loops. That something
like this might be the case has been claimed by
a recent study 
\cite{cka99,cka00}
of the properties of the magnetic centre symmetry 
at low and high $T$. This study
also suggests that the high-$T$ spatial string tension is only 
apparent, and that on appropriately large length scales the 
effective spatial string tension will vanish. 

The present work was originally motivated by these 
\cite{cka99,cka00}
unconventional claims. We set out to perform lattice Monte
Carlo calculations to determine the properties of
the short periodic vortices at high $T$ and to determine
whether the spatial string tension was indeed non-zero. The
claims made in 
\cite{cka99,cka00}
are strongest for D=2+1 gauge theories and so these are the
theories we consider. We work with the simplest SU(2) non-Abelian
gauge theory. Our work has, so far, involved a careful
reanalysis of some earlier, partially unpublished
calculations 
\cite{teper93kt}
of the spatial string tension as a function of $T$,
together with 
some simple probes of vortex properties. The results 
are striking and, if taken at face value, appear to disagree 
with the suggestion of
\cite{cka99,cka00}
concerning the behaviour of vortices and spatial 
confinement at high $T$.  Rather they provide
some quantitative evidence that vortices at low $T$ 
and (timelike) periodic vortices at high $T$ have a very
similar dynamics. Our work may also be viewed as complementing 
recent calculations 
\cite{kovacs00}
concerning the vortex free energy in D=3+1 low-$T$ SU(2)
gauge theories.

\section{SU(2) lattice gauge theory in 2+1 dimensions}
\label{sec_su2d3_review}

We work on a cubic lattice with gauge fields represented by 
SU(2) matrices, $U_\mu(n)$, on the links of the lattice.  
We use the standard plaquette action 
\be
S = \sum_p (1 - \frac{1}{2} Tr U_p)
\label{eqn_S_plaq}
\ee
where $U_p$ is the ordered product of link matrices around 
the plaquette labelled $p$. This action appears in the
partition function as 
\be
Z = \int \prod_l dU_l  \exp\{-\beta S\}
\label{eqn_Z}
\ee
where 
\be
\beta = \frac{4}{ag^2}.
\label{eqn_beta}
\ee
This relation holds in the continuum limit 
and we use it to define $g^2$ at arbitrary values of the
lattice spacing $a$. (Note that in 3 dimensions the gauge
coupling $g^2$ has dimensions of mass.)
It is clear from (\ref{eqn_beta}) that the continuum 
limit $a \to 0$ corresponds to $\beta \to \infty$.
This reflects the fact that the theory becomes 
free at short distances, just as in 4 dimensions.

To study the theory at finite temperature we use a
$L^2 \cdot L_t$  lattice, where $L_t$ is the number of  
sites in the periodic `temporal' direction, and
$L \to \infty$ in the thermodynamic limit. The temperature
of the system is given by
\be
aT = \frac{1}{L_t}.
\label{eqn_temp}
\ee

This lattice theory has been studied in detail in recent years
\cite{teper98}
and we summarise here some features that will be useful
for us later on.

\subsection{the confining phase at low $\bf T$}

The theory is linearly confining at low $T$. The numerical
evidence for this is no less clear-cut than in 3+1 dimensions.
The $T=0$ string tension 
\cite{teper98}
can be accurately parameterised as
\be
a\surd\sigma = \frac{1}{\beta}\biggl\{
1.337 + \frac{0.945}{\beta} + \frac{1.10}{\beta^2} \biggr\}
 \ \ \ \ \ \ \ : \beta \geq 3.0
\label{eqn_K_zeroT}
\ee
with an error of about $\pm 1 \%$. We can use (\ref{eqn_beta})
to convert (\ref{eqn_K_zeroT}) into an expression for the 
dimensionless ratio $\surd\sigma/g^2$ in terms of powers
of $ag^2$. When we need a value of $a\surd\sigma$, or of
$\surd\sigma/g^2$, at a value of $\beta$ at which direct
calculations
\cite{teper98}
do not exist, we shall use (\ref{eqn_K_zeroT}).

An alternative physical mass scale for the theory is
provided by the mass gap, $m_G$, which, here, turns out
to be the mass of the lightest $J^{PC} = 0^{++}$ state.
An accurate parameterisation for the ratio of $m_G$
to $\surd\sigma$
\cite{teper98}
is given by
\be
\frac{m_G}{\surd\sigma} = 
4.718 - \frac{1.083}{\beta^2}
 \ \ \ \ \ \ \ : \beta \geq 3.75
\label{eqn_mgap_zeroT}
\ee
with an error of $\leq \pm 1 \%$ for $\beta \geq 6$ and
somewhat larger at lower $\beta$. Using (\ref{eqn_beta})
and (\ref{eqn_K_zeroT}) we can convert (\ref{eqn_mgap_zeroT})
into an expression for $am_G$ or $m_G/g^2$.

\subsection{the deconfining transition at $\bf T=T_c$}

As the temperature of the system is increased, 
the theory undergoes a second order deconfining
phase transition at $T=T_c$. (See e.g. 
\cite{teper93tc,engels96}.) 
The values of $T_c$ in
\cite{teper93tc} 
can be accurately parameterised by 
\be
\frac{\beta}{4} \cdot aT_c \equiv
\frac{T_c}{g^2} = 
0.3875 \cdot \biggl(1-\frac{0.37}{\beta}\biggr)^{-1} 
\label{eqn_Tcrit}
\ee
which can be translated into a critical value of $L_t$
\be
L_t^c(\beta)  =
\frac{1}{1.55} (\beta - 0.37).
\label{eqn_Ltcrit}
\ee
We will use (\ref{eqn_Tcrit}) and (\ref{eqn_Ltcrit})
to express the value of $T=1/aL_t$ in units of $T_c$,
at the value of $\beta$ at which we are working.

\subsection{the deconfined phase at high $\bf T$}

In the deconfined phase, $T > T_c$, a Wilson
loop defined in a space-time plane will not
decay exponentially with its area. However
it has long been believed that purely spatial Wilson
loops do continue to show an area decay as in 
(\ref{eqn_wl_area}). That is to say, 
in the deconfined phase one has `spatial' linear confinement
with a corresponding non-zero string tension,
$\sigma(T)$, that will depend on the temperature.
A number of numerical studies have served to confirm
this expectation (see
\cite{bialas00} 
for a recent example), although it is probably 
fair to say that the effort put into this
question has been very much less than that
put into the study of confinement at $T=0$.
This perhaps leaves room for the suggestion
made in 
\cite{cka99,cka00}
that there is in fact no spatial confinement at high $T$
and the apparent spatial area decay is illusory.

Let us consider how this might be. The straightforward
possibility is that (accurate) numerical calculations 
have not been performed on distances that are larger than
the relevant dynamical length scale, and so the apparent
area decay simply reflects an approximate `accidental'
linearity in the spatial potential at small/intermediate 
distances. An `accidental' linearity is far from impossible.  
After all, the approximate linearity observed in the $T=0$ 
potential extends down to distances well below the 
characteristic distance scale $\sim 1 fm$ at which 
one would have expected a flux tube to have started forming.
Thus, the usual heavy-quark potential may be regarded 
as being `accidentally' linear at small distances.
What are `small distances' here? If $T$ is close to
$T_c$ then 
\be
\xi_c \equiv 1/T_c
\label{eqn_xic}
\ee
might be a relevant distance scale. At higher $T$ (see
\cite{altes96} 
for a review) we expect perturbation
theory (in the dimensionless coupling $g^2/T \ll 1$) to be
good, so that dynamically generated masses are $m^2 \sim g^2 T$,
and hence a typical dynamical length scale will be 
\be
\xi_T \equiv 1/gT^{1/2}. 
\label{eqn_xiT}
\ee
However this  ignores the infrared divergences in perturbation 
theory. For example, the Debye screening mass  $m_D$ is infinite 
at one loop. Fortunately direct numerical calculations of $m_D$ 
exist 
\cite{altes96} 
and these provide a reference scale 
\be
\xi_D \equiv 1/m_D.
\label{eqn_xiD}
\ee
We will consider these scales when we come to
discuss numerical results on the spatial string tension.

An alternative and more subtle possibility is provided
by the actual situation in QCD and similar theories where 
non-Abelian gauge fields are coupled to fundamental
matter fields
\cite{stringbreak}. 
Here direct numerical calculations of the potential
typically find it to be linearly rising
even beyond the distance at which one expects the string to 
break, and it requires a sophisticated mixing analysis 
to `see' the string breaking
\cite{stringbreak}. 
In such theories the breaking of the string is analogous
to the decay of a resonance. The string state exists,
but as an unstable state rather than as an asymptotic
state. In QCD the decay width is small and so the overlap 
of a good string operator onto the decay products will
be too small for the breaking to be visible in a 
calculation of limited accuracy. This scenario requires
an underlying confinement mechanism that is weakly
broken, as in QCD. If there is no confinement to start
with, as in weakly coupled D=3+1 QED, then this 
scenario is irrelevant
and there is no sign of any linearity in the potential.
While we are not aware of any underlying high-$T$ spatial 
confinement that would make the QCD example relevant
to the discussion of
\cite{cka99,cka00},
if it were to be so, then the fact that the string breaking is
a weak effect should make it easy to 
identify states into which it decays. For example, in 
the case of QCD it is into a light $q{\bar q}$ pair plus the
static sources. Then a mixing analysis 
\cite{stringbreak} 
could be performed to demonstrate the string breaking.

\section{The spatial string tension for $\bf T > T_c$}

In this section we shall analyse one particular calculation
\cite{teper93kt}
of the spatial string tension in the deconfined phase.
We shall look closely at the distance scales over which
we see linear confinement and ask whether these distances
are `large' or `small'. We shall also discuss the
values and behaviour of $\sigma(T)$, since, later 
on in the paper, we shall wish to compare our 
results with theoretical expressions that incorporate 
such a spatial string tension.

We shall need to discuss the results of 
\cite{teper93kt}
in some detail for two reasons. The first is that
some of what we present here has not been previously
published. The second is that 
\cite{teper93kt}
was not overtly interested in finite temperature physics 
and a careful translation of notation is needed.
The relevant calculations in
\cite{teper93kt}
were performed on $L_1 \times L_2 \times L_3$ lattices.
The Polyakov loops were along the 2 direction
and hence the periodic flux tubes onto which they projected
were of length $L_2$. The correlations were taken in the 
1 direction. The value of $L_3$ was varied and flux
tube masses were obtained for various values of $L_3$
and $L_2$. In
\cite{teper93kt}
$L_3$ was labelled $L_\perp$ and was referred to as a
transverse spatial direction: the point of the
calculation being to estimate the intrinsic width of
the confining flux tube. If we choose instead to call 
$L_t \equiv L_3$ 
the Euclidean time direction then what is calculated in
\cite{teper93kt}
is the mass, $M_p$, of a spatial confining flux tube of length 
$L \equiv L_2$ at a temperature of $aT = 1/L_t$. From this we 
can extract $\sigma(T)$ in the standard fashion:
\be
a M_p = a^2 \sigma(T) L - \frac{\pi}{6} \cdot \frac{1}{L} 
\label{eqn_poly_sig}
\ee
where the second term on the RHS is the universal string
correction for Polyakov loops. 
These calculations used Polyakov loops separated
in the 1-direction and individually smeared in a 2-3 plane.
The smearing, however, is only a device for improving the 
efficiency of the calculation. The lightest flux tube mass 
is the same as one gets with unsmeared Polyakov loops.
And this should give the same string tension as one gets with 
a large Wilson loop in the 1-2 plane, once the different
string corrections have been taken into account.

\subsection{linear confinement?}

\begin{figure}[p]
\begin{center}
\epsfig{figure=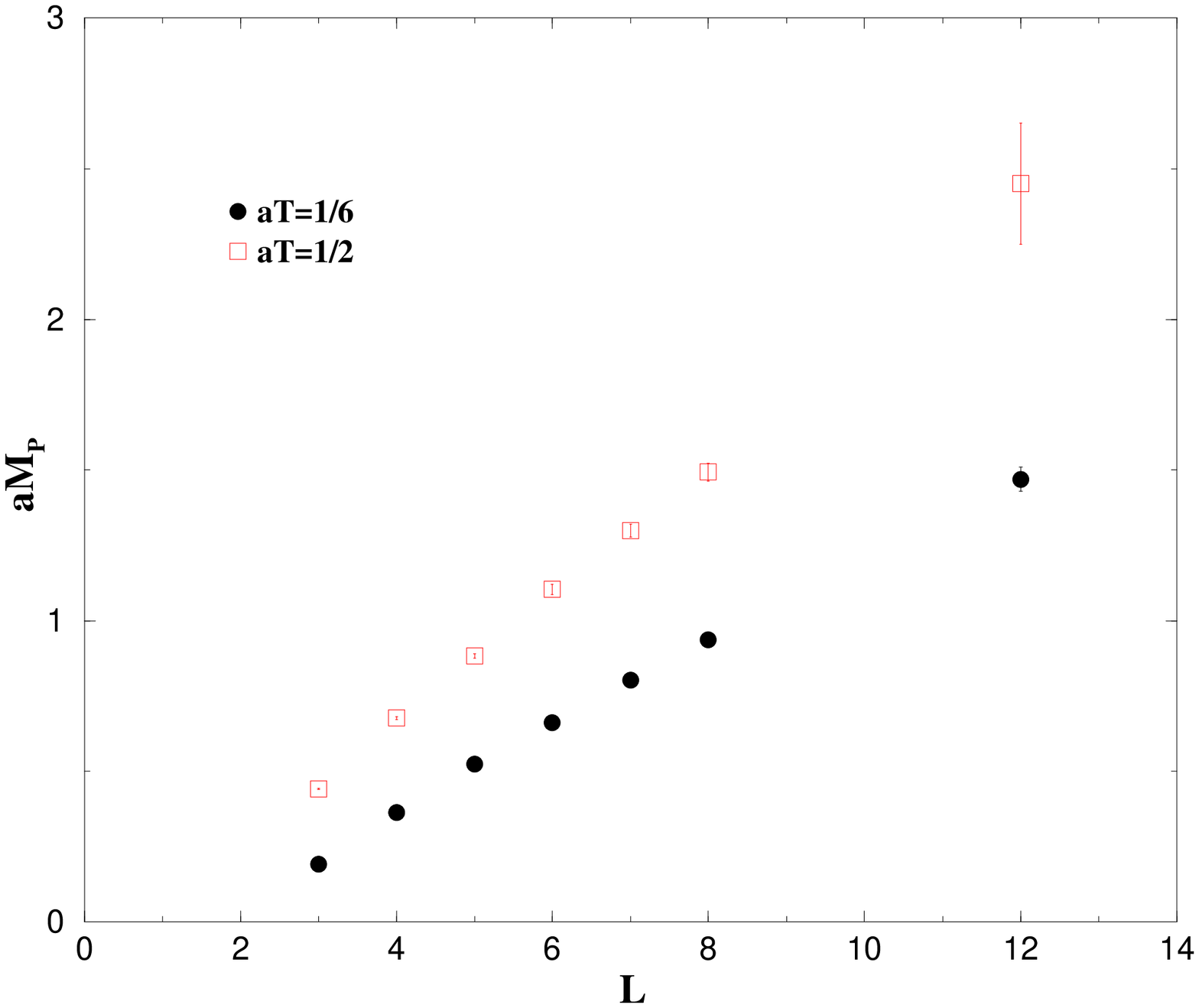, angle=270, width=15cm} 
\end{center}
\caption[]{\label{fig_mp_b45}
  { The mass of the spatial flux `tube' versus its
length for $aT = 1/2 > aT_c$ ($\Box$) and for 
$aT = 1/6 < aT_c$ ($\bullet$). All in lattice units at $\beta=4.5$.
}}
\end{figure}

The calculations in
\cite{teper93kt}
were performed at $\beta = 4.5$ and at $\beta = 9.0$.
What evidence do we have for linear confinement?
Linear confinement will, of course, be reflected in an 
approximate linear rise in the flux tube mass, $M_p$, 
with its length, $L$. In Fig.\ref{fig_mp_b45} we show how, 
at $\beta = 4.5$, this
mass varies as the flux tube length varies from $L = 3$
to $L = 12$. We show this for two different values
of $L_t$. The first is for $L_t =2$ which corresponds
to a temperature of $T = 1/2a = 1.33\times T_c$ (using
(\ref{eqn_temp}) and (\ref{eqn_Ltcrit})), which is well 
in the high-$T$ deconfined 
phase. The second is for $L_t =6$ which is well in
the low-$T$ confined phase. We see
an approximately linear rise throughout this range of $L$,
for both values of $T$.

Consider, in more detail, the $L_t=2$ high-$T$ calculation. 
The accurately calculated values are for $L \leq 8$. 
(The $L = 12$ mass is  so heavy that the errors become large.) 
Is a length of $8a$, at $\beta=4.5$ and at $T=1.33 T_c$,
large or small? 
Now the dynamical scales defined in 
(\ref{eqn_xic}), (\ref{eqn_xiT}) and (\ref{eqn_xiD})
are, here, $\xi_c \simeq 2.7a$, $\xi_T \simeq  1.5a $ and
$\xi_D \simeq 2.7a$. (For the last relation we 
have used the value $am_D = 0.373(5)$ which was calculated in
\cite{teper93tc}
for $\beta = 4.5$ and $aT = 1/2$, but which was not published.)
Compared to these length scales we can confidently claim to have 
seen linear spatial confinement over `large' distances.

Of course the $\beta = 4.5$ calculations correspond to a rather
coarse lattice spacing. At $\beta=9$, on the other hand, the 
lattice spacing is well into the weak coupling scaling region 
\cite{teper98}
and here there exist calculations
\cite{teper93kt}
of flux tube masses for two lengths, $L = 8$ and $L = 16$. One 
again observes an approximate linear rise at both low and high $T$.
For example, at  $L_t=4$ (where $T \simeq 1.4 T_c$)
one finds $aM_p(L=8a) = 0.292(3)$ and $aM_p(L=16a) = 0.677(6)$.
If we evaluate the previously discussed high-$T$ dynamical length 
scales at this $\beta$ and $L_t$, we find $\xi_c  \simeq 5.6a$,
$\xi_T  \simeq  3a$ and $\xi_D  \sim 5a$. (For the last relation 
we do not have a directly calculated value of $m_D$ and have 
instead interpolated between `neighbouring' values calculated in
\cite{altes96}
and unpublished work from
\cite{teper93tc}.)
So here too we can claim to have seen linear spatial 
confinement on reasonably `large' distance scales.

From the above we feel able to say that the spatial linear
confinement we observe is not an accidental linearity
obtained over small distance scales. On the other hand,
the possibility of a weak string-breaking is something we
cannot address without a more specific scenario for
the string-breaking dynamics.

\subsection{$\bf \sigma(T)$}

\begin{figure}[p]
\begin{center}
\epsfig{figure=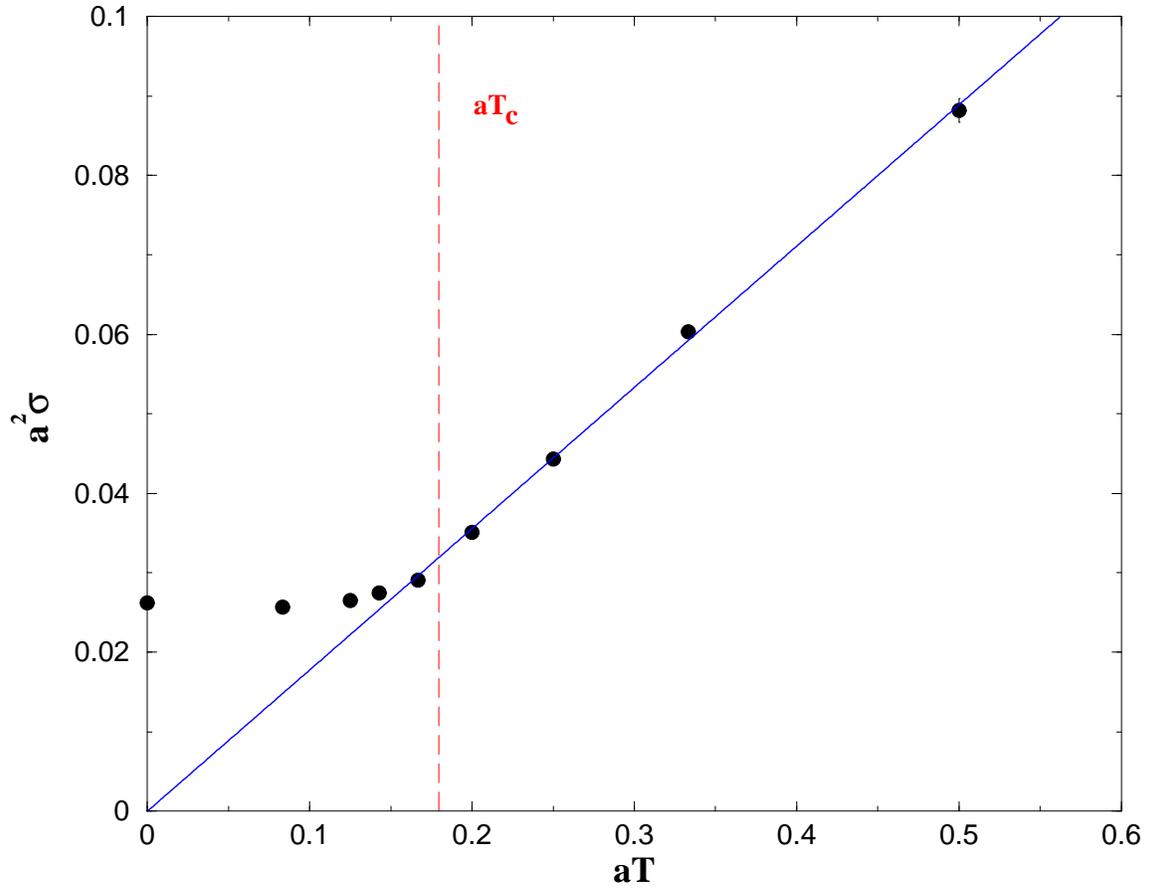, angle=270, width=15cm} 
\end{center}
\caption[]{\label{fig_sigT_b90}
  {The spatial string tension as a function
of $aT$ at $\beta=9$. The deconfining temperature, $aT_c$,
is indicated.
}}
\end{figure}

We take from 
\cite{teper93kt}
the values of the flux tube mass, $M_p$, at $\beta=9$,
for various values of the temperature $aT=1/L_t$
and for a flux tube length of $L=16a$. We then
use (\ref{eqn_poly_sig}) to extract the corresponding
string tensions, $\sigma(T)$, which we plot in 
Fig.\ref{fig_sigT_b90}.
We see that $\sigma(T) \propto T$ for $T > T_c$,
just as one expects from the simplest dimensional
reduction arguments. (This observation, in a
different notation, was already made in 
\cite{teper93kt}
and a simple dynamical reason for it was given there.)
More specifically we find:
\be
\sigma (T) \simeq 0.40 g^2 T \simeq T_c T 
 \ \ \ \ \ \ \ \ : \ T>T_c .
\label{eqn_sigmaT}
\ee
At $\beta=4.5$ only the $L_t = 2$ calculation is in the 
deconfined phase, so we cannot perform a similar
calculation there. However if we assume a linear
rise in $T$ then we obtain a coefficient of $\simeq 0.44$
rather than the $0.40$ in (\ref{eqn_sigmaT}). The size of
the difference is consistent with being a $O(1/\beta)$ 
lattice correction and, if so, this implies that
the continuum value of the coefficient will be $\simeq 0.36$.
Clearly, a more detailed calculation to control the continuum
limit would be useful.
 
In a later section we shall need a particular value of the 
string tension:
\be
a^2 \sigma (T= 1/4a) = 0.0443(4) \ \ \ \ \ \ : \beta = 9.0.
\label{eqn_sigmaT90}
\ee
\section{Vortices and boundary conditions}
\label{sec_vor_twist}

\subsection{centre vortices on an infinite lattice}

Consider an arbitrary gauge field, $\{U_\mu(n) \}$, on an infinite
lattice. Let the sites be labelled by integers $(n_x,n_y,n_t)$.
We can add an ultraviolet vortex to this field by the following
transformation on the fields, $U \to U^\prime$. Choose some
particular integers $n^v_x,n^v_y$ which will label the
spatial plaquette threaded by the vortex line. We take 
$U^\prime_l = U_l$ for all links $l$ except that
\be 
U^\prime_y(n_x,n^v_y,n_t) = - U_y(n_x,n^v_y,n_t)
\ \ \  \ \  {\rm for} \  n_x \leq n^v_x \ {\rm and} \ \forall n_t.
\label{eqn_uv_vortex}
\ee
It is easy to see that this transformation flips the sign of
the spatial $(x,y)$ plaquettes emanating from the sites
$(n^v_x,n^v_y,n_t)$, for all values of $n_t$, but that
it leaves all other plaquettes unchanged. (Since $-1$ is an element 
of the centre of the group and so commutes with all $U_l$.)
Moreover if we consider an arbitrary spatial Wilson loop,
it is either unchanged or its sign is flipped. The latter
happens if and only if the Wilson loop encircles a plaquette
whose sign has been flipped. This has all the expected
properties of a vortex. It is an ultraviolet vortex
because the vortex core, where the action has changed,
is a string of elementary plaquettes. The change of
sign of these plaquettes means that their average action
has increased; and that the action density along the string
will diverge in the continuum limit. The associated 
$Z_2$ gauge transformation has also been localised
onto an ultraviolet length scale: a single link. 
Such a vortex may be suppressed by its action, but
the measure of the transformed field is clearly the 
same as that of the original field. 

In the above construction there is a half-plane of 
links that acquires a minus sign. This half-plane can
be smoothly deformed while keeping the edges fixed.
On an infinite lattice one edge is invisibly far,
and we can deform a set of $\mu=y$ negative links 
heading off to $n_x = -\infty$ to, say, a set of $\mu=x$
negative links heading off to $n_y = -\infty$. On
a finite lattice this will no longer be possible.

It is also clear that we can introduce any number of 
these ultraviolet vortices,
with vortex lines at any positions. By allowing $n^v_x, n^v_y$
to vary smoothly with $n_t$ we can introduce vortices whose  
world lines are not straight. A suitable pair of such
vortices can produce a vortex line that is a closed loop
in space-time. Thus an ensemble of ultraviolet vortices
may consist of closed loops of arbitrary sizes and shapes.

When quantum fluctuations are included, we expect that if 
vortices do exist then the core will acquire a size $O(\xi)$
where $\xi$ is an appropriate dynamical length scale. (And
that in any case the gauge transformation will typically  
be spread over distances greater than one link.) 
However any such field can be reached
by a smooth deformation of the ultraviolet vortex we have 
defined above, and so, for the `kinematic' part of our 
discussion, it suffices to focus on the latter.

\subsection{twisted and periodic boundary conditions}

From now on we shall be in a finite spatial volume.
Suppose that we are on a torus so that the gauge
potentials are periodic in $x$ and $y$, with 
periods $L_x$ and $L_y$ in lattice units. 
If a gauge field $\{U_l\}$ is periodic then it
is obvious that when we add an ultraviolet vortex, 
as in (\ref{eqn_uv_vortex}), then the transformed
field $\{U^\prime_l\}$ is not periodic. However
we can add two such vortices, located at say
$n^{v1}_x,n^{v1}_y$ and $n^{v2}_x,n^{v2}_y$,
since we can deform the two strings of
flipped links so that they join the two vortex
cores and there is no string of flipped links 
heading off to $n_x = - \infty$. A Wilson loop
that encircles both vortices will not flip sign
but one that encircles only one, will. Thus
a periodic gauge field possesses an even number 
of such vortices. 

To obtain fields with an odd number of vortices
we need to impose twisted boundary conditions.
We shall do this in the usual way, keeping
the boundary conditions periodic but transforming the
plaquette action in (\ref{eqn_S_plaq}) as follows 
(see
\cite{altes96}
for a review): 
\be
S_{tw} = 
\sum_{p\not= p^\prime} (1 - \frac{1}{2} Tr U_p)
+
\sum_{p^\prime} (1 + \frac{1}{2} Tr U_{p^\prime})
\label{eqn_S_twist}
\ee
where the set of plaquettes $p^\prime$ is specified
to be the set of $(x,y)$ plaquettes emanating
from some specific spatial 
coordinates $n^{tw}_x,n^{tw}_y$ at all values of
$n_t$. That is to say we flip the sign of a particular
string of spatial plaquettes in the action, the string
closing on itself through the periodic temporal boundary. 
That this is equivalent to the usual plaquette action with
twisted boundary conditions can be shown by making
a singular centre gauge transformation followed by
a suitable redefinition of field variables.

Consider a field $\{U\}$ that is typical of those that one obtains
with the usual plaquette action
(\ref{eqn_S_plaq}). Now consider the same field, 
but with respect to the twisted action. The action density is 
unchanged, except along the line of twisted plaquettes
where its sign is flipped. This is just what we would expect
if we had added an (ultraviolet) vortex line so as to thread 
this line of plaquettes. That is to say, we appear to have added
a single vortex to the field $\{U\}$, something that we were
previously unable to do.

This interpretation becomes clearer if the vortex is not 
superimposed upon the twisted plaquettes, but is added
to  $\{U\}$ so that it threads some other line of plaquettes, 
labelled by $n^v_x,n^v_y$. We do this, as usual, by multiplying 
by $-1$ a string of links running off from $n^v_x,n^v_y$.
Now, instead of running this string to the boundary (which
would lead to problems with the periodic boundary conditions) 
let us run it up to $n^{tw}_x,n^{tw}_y$ where the twisted 
plaquette lies. This will flip the value of the twisted 
plaquettes. So now the action density of this field with
a twist is the same as that of the original field $U$
without a twist, even at the location of the twist, except
that the plaquettes at the location of the vortex core have 
been flipped. This is exactly what we got when we added
a single vortex to a field in an infinite volume.
Moreover, any Wilson loop enclosing the vortex, but
not the twist, will acquire a factor of $-1$ relative to
its value in the field $U$. We may think of the twist
as adding a vortex source to the system. Alternatively
we note that if we choose to trade the twisted action 
for twisted boundary conditions, then we obtain a string 
of flipped links from the vortex to the twist and then on to
the boundary. But now this presents no problem because if the field
$\{U\}$ was periodic, the field with a vortex added satisfies
the twisted boundary conditions. Thus the ensemble of fields 
obtained with a twisted action is just like the ensemble from
the simple plaquette action, but with a vortex added or subtracted.
If periodic field have even vorticity then twisted fields
have odd vorticity.

\subsection{vortex free energy, $\bf F_v$}

Consider a system at temperature $T=1/aL_t$. The partition
function and free energy are related by
\be
\exp\biggl\{ -\frac{F}{T} \biggr\} = Z.
\label{eqn_F_defn}
\ee
Let $F_{tw}$ be the free energy of the system with a twist, 
and $F_{nt}$ of the identical sytem with no twist. We define
the vortex free energy to be the difference
\be
F_v = F_{tw} - F_{nt}.
\label{eqn_Fv_defn}
\ee
Note that this will not be the free energy of a single
vortex, unless the usual vacuum contains no vortices. 
It is the free energy difference between a system 
with an odd number of vortices, and a system with 
an even number.

From (\ref{eqn_Fv_defn}), (\ref{eqn_F_defn}) and 
(\ref{eqn_Z}) we readily obtain
\be
\frac{\partial}{\partial\beta} \biggl( \frac{F_v}{T} \biggr) 
=
\frac{g^2}{4L_t} \cdot
\frac{\partial}{\partial T} \biggl( \frac{F_v}{T} \biggr) 
=
\langle S_{tw} \rangle - \langle S_{nt} \rangle
\label{eqn_derivFv}
\ee
where $S_{tw}$ is defined in (\ref{eqn_S_twist}), $S_{nt}$
is the usual untwisted action defined in (\ref{eqn_S_plaq}),
and we transform the derivative with respect to $\beta$ 
into one with respect to $T$ using the fact that the former
is performed at fixed $L_t$. 

If $F_v \not= 0$ in the thermodynamic limit ($L_x,L_y \to \infty$)
then the vortices certainly do not condense in the vacuum
and will not contribute to a linearly confining `spatial potential'.
This is known to occur for vortices in a space-time plane at
high $T$
\cite{altes96}.
In this paper we calculate not $F_v$ itself, but rather
$\partial/\partial T ( F_v / T ) $ using (\ref{eqn_derivFv}).
We shall assume that if the latter is zero, then so is
the former (since we see no reason for $F_v$ to be
exactly linear in $T$). A direct calculation of $F_v$
is possible but less straightforward
\cite{kovacs00}.
\subsection{action of a single vortex}

A single ultraviolet vortex imposed on a gauge field with 
$U_l = 1 \ \forall l$ will clearly have an action
\be
 S_v = 2 L_t = \frac{2}{aT}.
\ee
If the vortex is fat, so that the $\pi$ units of flux are
smoothly spread over $A_v$ plaquettes, then 
\be
 S_v = \sum_p (1-\cos\theta_p) 
\simeq \frac{1}{2} \frac{\pi^2}{A_v} L_t.
\label{eqn_Svortex_fat}
\ee
If we are working with fluctuating gauge fields at some
value of $\beta$ then
in general we expect the vortex action to be modified by
a lattice renormalisation factor:
\be
S_v(\beta) = S_v(\infty) \cdot Z_v(\beta). 
\label{eqn_Svortex_beta}
\ee

We note that if, for example, 
$A_v \sim (\xi_D/a)^2 = 1/(a^2 \cdot g^2 T)$
then $\beta S_v$ is independent of $a$ or $T$.
On the other hand, if the vortex size is not
stabilised by quantum fluctuations, it will
spread over the whole spatial volume, $L^2$, so that
\be
S_v \simeq 
\frac{\pi^2}{2}\cdot \frac{L_t}{L^2} \cdot Z_v(\beta)
\stackrel{L\to\infty}\longrightarrow 0.
\label{eqn_Svortex_spinwave}
\ee
In this case the vortex merges into the Gaussian 
spin-wave contribution and ceases to be a useful degree
of freedom. This is indeed what occurs in  U(1) lattice
gauge theories: the flux is conserved and hence
comes in closed loops, but these do not come in
vortices of finite area. Although Wilson loops are
still disordered, the effect is Coulombic rather than
linearly confining.

\subsection{domain walls and spacelike vortices at high T}

To give substance to the above discussion of vortices,
it would be useful to identify a situation where the vortex 
becomes a real and easily visible excitation, just like a
magnetic monopole in the Higgs phase of the Georgi-Glashow
model. It turns out that such a situation exists
\cite{altes96}.
Consider a spatial volume $L \times L_\perp$ at a temperature
$T$ close to zero. Suppose also that $L$ is large. If one 
decreases $L_\perp$ from large values, one finds that there
is a phase transition at some $L_\perp = L^c_\perp(\beta)$.
If in addition the system possesses a spatial twist, then,
for $L_\perp \ll L^c_\perp(\beta)$, one finds that there is a 
single vortex whose world line extends through time, and
which close upon itself through the periodic temporal boundary 
conditions. Moreover the core of this vortex extends right
across the short $L_\perp$ direction. In the longer
$L$ direction it has a size $\propto \surd(\beta L_\perp)$.
As $L_\perp \downarrow$ the vortex action density grows
as $S_v \propto 1/L^{\frac{3}{2}}_\perp$ and the probability
of any virtual vortex quantum fluctuation rapidly vanishes.
Thus, the special spatial geometry drives the system into 
an effective vortex Higgs phase, in which the vortex appears 
`squeezed' by the short spatial direction. Apart from this 
distortion, it is precisely a vortex of the kind we have been
discussing. 
 
In
\cite{altes96}
the context of these objects was (apparently) quite different,
and we need to provide a brief translation. Consider the
system described in the previous paragraph. 
Rename the time direction to be a spatial one and 
the short $L_\perp$ direction to be the temporal one. Then
the phase transition of the previous paragraph is simply 
the deconfining transition, $T_c=1/(aL^c_\perp)$,
and the squashed vortex is the `domain wall'
separating two high-$T$ $Z_2$ vacua. (Or, more precisely,
it is the world sheet of the string-like boundary that
separates the two  $Z_2$ vacua.) The Polyakov loops
on either side of the vortex have a relative minus sign:
just as one would expect for a centre vortex. It was
briefly pointed out in 
\cite{altes96}
that these domain walls have a simple interpretation as 
real vortices in a space-time plane. Renaming axes it
becomes a spatial vortex as discussed in the previous paragraph.

\section{Vortices and confinement}
\label{sec_vor_conf}

In this section we shall assume that centre vortices exist
at high $T$ and that they are the fluctuations  which produce 
the spatial string tension. We assume (reasonably) that the
relevant vortices are the short ones whose world lines close through
the temporal boundary and whose length is therefore $1/T$.
These vortices will have a typical area $A_v$, and a corresponding
action, $S_v$, as in (\ref{eqn_Svortex_fat},\ref{eqn_Svortex_beta}).
Note that we are talking here of the vortices in the Euclidean field
configurations, rather than the vortices that may appear in the
eigenstates of the Hamiltonian. Thus $A_v$ may depend on $T$. 

In the following calculations we shall assume that the 
vortices are dilute, since it makes the argument simpler. 
Whether this is actually so is something we
shall return to in the next Section when we discuss our
numerical results. As long as vortices  are uncorrelated at 
sufficiently large distances, our conclusions will not
be qualitatively altered by vortex correlations which,
to a first approximation, can be absorbed into a
renormalisation of the effective vortex action. 

We shall first show how large Wilson loops acquire
an area decay in the presence of such a gas of vortices.
We shall ask what the observed $T$ dependence of 
$\sigma(T)$ teaches us about the properties of these
vortices. We shall then show how the average action
with twist approaches the periodic action as the
volume$\to \infty$, and how this
approach is directly related to the string tension 
generated by the gas of vortices. This parallels
the well-known results obtained in 
\cite{thooft}
for $T=0$.

\subsection{vortices and the spatial string tension}

Consider a spatial Wilson loop defined along a curve $c$ that 
bounds a minimal area $\cal A$ (in lattice units). 
Let us calculate its value within
a simple gas of vortices. Suppose that the area $\cal A$ is 
pierced by $n_v$ vortices. Assume they are uncorrelated so
that the probability is given by a simple Poisson formula
\be
P(n_v;{\cal A}) = 
{ {(\mu_0 {\cal A})^{n_v}} \over {n_v!} } \cdot 
e^{-\mu_0 {\cal A}}
\label{eqn_poisson}
\ee
where $\mu_0$ is the probability of a vortex piercing a
unit area. Now each vortex contributes a factor $-1$ to
the value of the Wilson loop, so in this simple picture
\begin{eqnarray}
\langle W_c \rangle = 
\langle Tr \ e^{\int_c A \cdot ds} \rangle & = &
\sum_{n_v} (-1)^{n_v} P(n_v;{\cal A}) \nonumber \\
& = & \sum_{n_v \ even} P(n_v;{\cal A}) 
- \sum_{n_v \ odd} P(n_v;{\cal A}) 
 \nonumber \\
& = & e^{-2\mu_0 {\cal A}}, 
\label{eqn_WL_poisson}
\end{eqnarray}
where the last relation follows from using (\ref{eqn_poisson}).
That is to say, the gas of vortices produces linear confinement
with a string tension $a^2\sigma = 2 \mu_0$.  
 
Since we are at high $T$ we expect the relevant vortex world lines 
to be closed through the short temporal boundary. If their
cross-sectional area is $A_v$ (in lattice units) then we 
expect $\mu_0$ in (\ref{eqn_WL_poisson}) to contain a factor 
$\sim 1/A_v$ as well as a factor containing the action
(\ref{eqn_Svortex_fat},\ref{eqn_Svortex_beta}):
\begin{eqnarray}
a^2 \sigma(T)  = 2\mu_0 
& \sim & \frac{2}{A_v} e^{-\beta S_v} 
\label{eqn_mu_poisson1} \\
& \simeq &
\frac{2}{A_v} e^{ -
\frac{\beta \pi^2}{2 A_v} \cdot \frac{1}{aT} \cdot Z_v(\beta)}.
\label{eqn_mu_poisson2}
\end{eqnarray}
One uncertainty here is in the integration over spatial
translations of the vortex. This will produce, in addition to 
the factor $1/A_v$, a factor, presumably $O(1)$, that depends 
on the details of the collective co-ordinate 
calculation. There will also be a factor from the fluctuations 
around a vortex. As we remarked earlier, a natural length scale
at high $T$ is provided by $\xi_D = 1/\sqrt{g^2 T}$. If we assume
that $A_v \propto (\xi_D/a)^2$ then we find from (\ref{eqn_mu_poisson2})
and (\ref{eqn_WL_poisson}) that the string tension is
\be
a^2 \sigma(T) = 2\mu_0 = c a^2 g^2 T
\label{eqn_mu_sigma}
\ee
where $c$ may have a weak $\beta$ dependence from the $Z_v(\beta)$ 
factor. Thus we naturally obtain a spatial string tension that
increases linearly with the temperature (as one would expect from
naive dimensional reduction). Conversely the observation
that $\sigma(T) \propto T$ implies that $A_v \propto 1/T$ and
hence that the weighting $\exp(-\beta S_v)$ is independent of $T$.

\subsection{$\bf \Delta S$ and vortices}

What do we expect for the action difference
\be
\Delta S \equiv \langle S_{tw} \rangle - \langle S_{nt} \rangle
\label{eqn_deltaS}
\ee
that appears in (\ref{eqn_derivFv})? The only difference between
having twisted and periodic boundary conditions on our 
$L \times L$ spatial volume is that these constrain us to having an 
odd or even number of vortices respectively. We assume
that in both cases we have a Poisson distribution, just as in
(\ref{eqn_poisson}) with ${\cal A}=L^2$. Thus the probability of
$n_v$ vortices is 
\be
P(n_v; L^2) = \frac{1}{\cal N} \cdot
{ {(\mu_0 L^2)^{n_v}} \over {n_v!} }
\label{eqn_poisson_spatial}
\ee
where the normalisation factors in the two cases are
\be
{\cal N}_{tw} 
=
\sum_{n_v \ odd} { {(\mu_0 L^2)^{n_v}} \over {n_v!} }
=
\sinh (\mu_0 L^2)
\label{eqn_poisson_normtw}
\ee
\be
{\cal N}_{nt} 
=
\sum_{n_v \ even} { {(\mu_0 L^2)^{n_v}} \over {n_v!} }
=
\cosh (\mu_0 L^2).
\label{eqn_poisson_normnt}
\ee
Let $s_0$ be the effective action of a single vortex.
Then it is a simple calculation to obtain
\begin{eqnarray}
\Delta S
& = &
\frac{1}{{\cal N}_{tw}} \sum_{n_v \ odd} n_v s_0
{ {(\mu_0 L^2)^{n_v}} \over {n_v!} }
-
\frac{1}{{\cal N}_{nt}} \sum_{n_v \ even} n_v s_0
{ {(\mu_0 L^2)^{n_v}} \over {n_v!} }  \nonumber \\
& = &
s_0 \cdot \mu \frac{\partial}{\partial\mu} \ln(\tanh \mu)
\nonumber \\
& = &
2s_0 \cdot a^2\sigma L^2 \cdot
{{e^{-a^2\sigma L^2}} \over {1-e^{-2a^2\sigma L^2}}}
\label{eqn_deltaS_poisson}
\end{eqnarray}
where in the second line we use the shorthand notation
$\mu \equiv \mu_0 L^2$ and at the final step we have used 
$2\mu_0 = a^2\sigma(T)$ from (\ref{eqn_mu_sigma}).

In (\ref{eqn_deltaS_poisson}) the overall factor of 
$\exp\{-a^2\sigma L^2\}$ is no surprise in a linearly confining
theory: it arises from the fact that we are subtracting
the partition function for an odd number of vortices from
that for an even number, just as in (\ref{eqn_WL_poisson}). 
The prefactor of $\sigma L^2$ arises from the 
average number of vortices in the $L^2$ volume; both 
$\langle S_{tw} \rangle$ and $\langle S_{nt} \rangle$ are 
proportional to it for large volumes; and hence so is 
their difference, $\Delta S$. 
For small enough spatial volumes we expect the periodic
volume to have no vortices and the twisted volume to have just 
one vortex. We observe that (\ref{eqn_deltaS_poisson}) embodies
this expectation: in the limit $L^2 \to 0$, one obtains
$\Delta S \to s_0$, since the factor in the denominator
cancels the $\sigma L^2$ prefactor in the numerator.  
Of course, for realistic vortices, the action will start
to vary once $L^2 < A_v$ and this simple formula will
break down.

\section{Calculations of $\bf \Delta S$}
\label{sec_results}

In this section we present the results of our lattice Monte 
Carlo calculations of  $\Delta S$. This is defined 
(\ref{eqn_deltaS}) as the difference of the average total 
action on lattices with and without a twist:
$\Delta S \equiv 
\langle S_{tw} \rangle -\langle S_{nt} \rangle$.
It is related (\ref{eqn_derivFv}) to the derivative 
with respect to  $T$ of the vortex free energy $F_v$.
In the low $T$ deconfining phase we expect that
$\lim_{L \to \infty} F_v = 0$ and
hence $\lim_{L \to \infty} \Delta S = 0$. If 
we have spatial confinement at high $T$ then we
expect this to hold there as well. At low $T$
there are specific expectations about how this
limit is approached
\cite{thooft}.
At high $T$ we can test the possibilities described in 
the previous section.

Our individual calculations of $\langle S_{tw} \rangle$
and $\langle S_{nt} \rangle$ consist, typically, of  
$O(10^6-10^7)$ interweaved heat bath and over-relaxed sweeps. 
We shall begin with a calculation of $\Delta S$ in
large spatial volumes and for temperatures ranging
from well below $T_c$ to well above. We shall then
present a finite volume study of $\Delta S$ for
$T > T_c$. We shall finish with a similar finite
volume study of the system in the low $T$ confining phase.

\subsection{$\bf \Delta S$ on large spatial volumes}
\label{subsec_dSlargeV_results}

\begin{figure}[p]
\begin{center}
\epsfig{figure=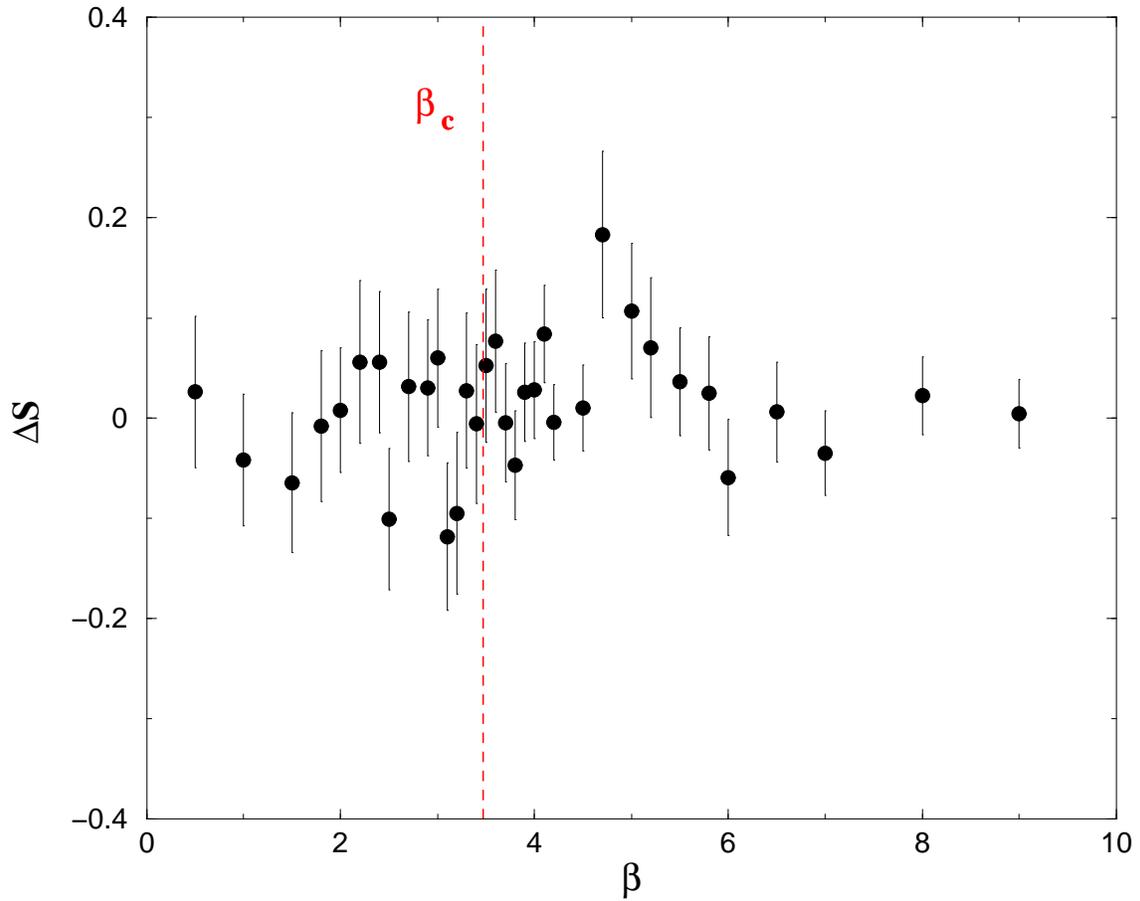, angle=270, width=15cm} 
\end{center}
\caption[]{\label{fig_Sdiff_scan}
  {The difference of twisted and untwisted actions 
on $L_t=2$ lattices. 
$T$ increases as $\beta$ increases, and the
system deconfines at $\beta_c$ as indicated.
The spatial volumes are large in physical units,
except very close to $\beta = \beta_c$ where
there is a diverging correlation length.
}}
\end{figure}

Our first calculation is of 
$\Delta S \equiv \langle S_{tw} \rangle - \langle S_{nt} \rangle$
in the limit of large volumes and for temperatures ranging from 
values that are well within the confining phase, to values much larger 
than $T_c$. The vortex free energy (\ref{eqn_derivFv}), and hence 
$\Delta S$,
should certainly be zero for $T < T_c$, so the interesting
question is whether there is a change for $T > T_c$.
In Fig.\ref{fig_Sdiff_scan} we show our results.
As is clear, these are compatible with $\Delta S = 0$ over 
our whole range of $\beta$. However, to assess the significance
of this result, there are some questions that need to be answered. 
First, what is the range of $T$? Then, how large is the spatial 
volume? Finally, and most importantly, are the errors on these 
values of  $\Delta S$ small compared to the non-zero value 
one is attempting to discriminate against?

First, the range of $T$. On our $L_t =2$ lattice deconfinement
occurs for $\beta=\beta_c=3.47$ (see (\ref{eqn_Ltcrit})). The range of 
$\beta$ values for this calculation is $\beta \in [0.5,9.0]$. 
Using (\ref{eqn_beta},\ref{eqn_temp}) this translates into
$0.15 T_c \leq T \leq 2.6 T_c$. (Note that because of large lattice
corrections at small $\beta$, the lower limit has to
be interpreted with care.) 

Is the volume large? We use $L=12$ for $\beta < 2.5$,
$L=16$ for  $2.5 \leq \beta < 4.5$, and $L=24$ for 
$\beta \geq 4.5$. First consider $T < T_c$, i.e. $\beta < 3.47$. 
Here a relevant physical
length scale is provided by $\xi_\sigma \equiv 1/\sigma(T=0)$. 
In these units 
\cite{teper98}
we have $aL/\xi_\sigma > 7.8$ for $T < T_c$. As $T \to T_c$ there 
is a correlation length, $\xi_p$, that diverges (the transition
is second order) and so there will be some region of $\beta$ 
near $\beta_c$ where our volume will become too small to
accommodate  $\xi_p$. Such finite volume effects have been
studied in detail in
\cite{teper93tc}
and one finds that they become significant on a $16^2 2$ lattice
only for $\beta > 3.15$ where $\xi_p > 7$. Going now to $T > T_c$
we note that there is again an interval close to $T_c$ where
the correlation length diverges. We do not have direct
calculations of finite size effects here, but we assume that
they are negligible for $\beta \geq 3.7$ since
$\xi_p(\beta=3.7) \simeq 3$ and $\xi_D(\beta=3.7) \simeq 6$
\cite{teper93tc}.
In the high-$T$ range our lattice sizes range from 
$aL \sim 4.5\xi_D$ at $\beta = 4.0$ to
$aL \sim 9\xi_D$ at $\beta = 9.0$. Thus, apart
from a window of temperatures close to $T_c$, the 
volumes we are using are indeed large. (In fact there
is some reason to believe that even within this window 
the quantities we are interested in will be little affected.
For example, the calculated values of the spatial string
tension, $\sigma(T)$, show no effects as $T$ passes near
$T_c$. And we see no sign of larger fluctuations near
$T_c$ in either $\sigma(T)$ or $\Delta S$.) We shall later
perform some detailed finite volume studies that will
confirm this.

How significant is our result that $\Delta S$ is compatible 
with zero? The extra action of a single ultraviolet vortex 
would arise from the change in sign of a single plaquette
in each time-slice, and so it would
contribute $\delta S_v = L_t \langle Tr U_p \rangle$. This
increases from 
$\delta S_v \simeq 2.5$ at $\beta =3.0$ to
$\delta S_v \simeq 3.5$ at $\beta =9.0$. This is certainly
excluded by Fig.\ref{fig_Sdiff_scan}. The action of
a more realistic `fat' vortex will depend on the area,
as in (\ref{eqn_Svortex_fat}) and (\ref{eqn_Svortex_beta}).
Our results constrain the diameter of such a vortex to
be $d_v > 10a$. Compared to the natural length scales
in the problem (as discussed above) this is very large,
and thus we believe that our results for $\Delta S$
strongly discriminate against the twisted system differing 
from the untwisted one by the presence of a single
{\it extra} vortex.

\subsection{$\bf \Delta S$ as a function  of $\bf L$  for $\bf T > T_c$}
\label{subsec_dShighT_results}

We have seen that $\lim_{L \to \infty} \Delta S = 0$.
How is this limit approached? Does it point to 
a gas of confining vortices, and hence a behaviour
as in (\ref{eqn_deltaS_poisson})? Or is it just that
vortices do not exist; for example as in 
(\ref{eqn_Svortex_spinwave}) because their
size is not stabilised?

\begin{figure}[p]
\begin{center}
\epsfig{figure=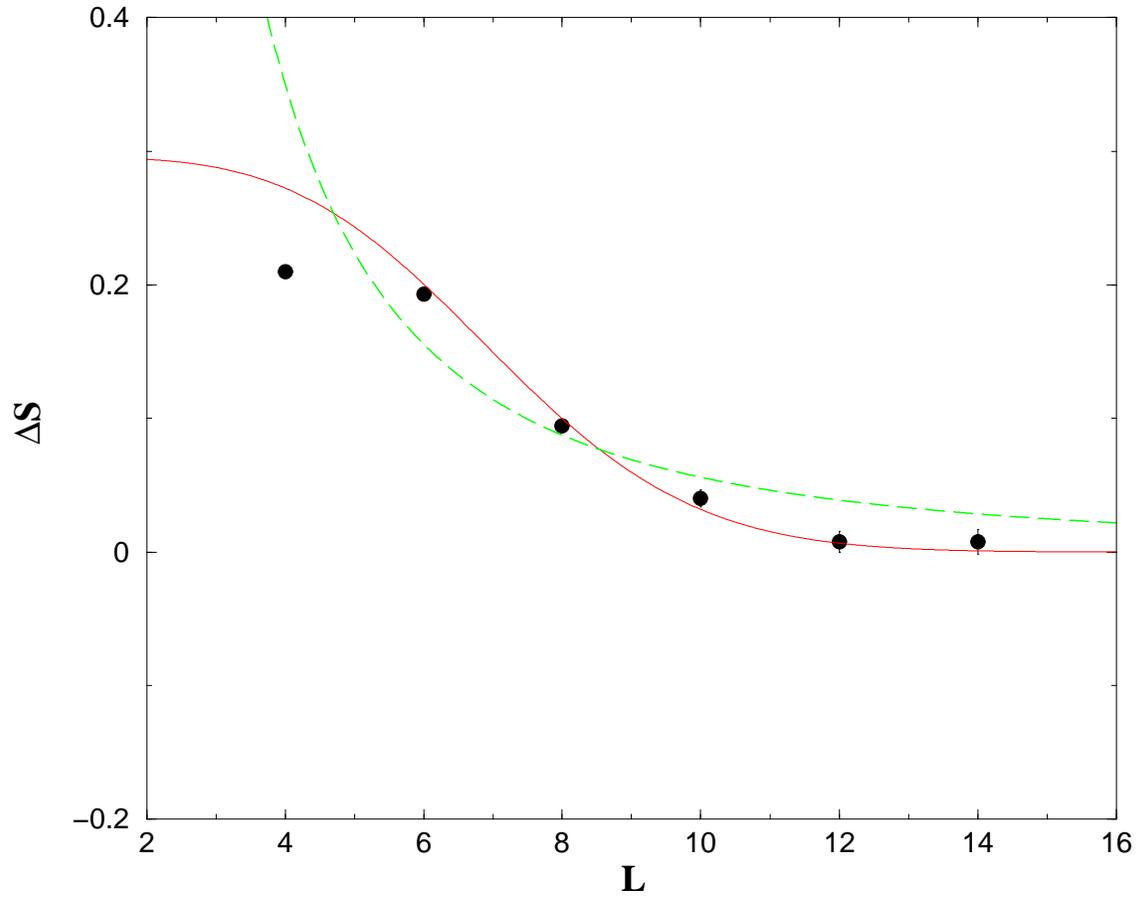, angle=270, width=15cm} 
\end{center}
\caption[]{\label{fig_Sdiff_deconf4}
  {The difference of twisted and untwisted actions 
at $T=1/4a > T_c$ at $\beta = 9$, for various spatial sizes.
Curves correspond to best fits for a dilute vortex gas
(solid line) and for a single unstable `vortex' (dashed line).
}}
\end{figure}

To address this question we have calculated $\Delta S$
as a function of the spatial lattice size, $L$, at $\beta=9$
and for $L_t = 4$ which, at this $\beta$, corresponds 
(\ref{eqn_Ltcrit}) to $T \simeq 1.4 T_c$. We show our
results in Fig.\ref{fig_Sdiff_deconf4}. We observe that
on small volumes $\Delta S$ does indeed deviate 
substantially from zero. Moreover, we see that the
variation is not compatible with the 
$\Delta S \propto 1/L^2$ behaviour expected
(\ref{eqn_Svortex_spinwave}) if the
`vortex' spreads to fill the whole spatial volume.
By contrast it is well fitted by the expression
(\ref{eqn_deltaS_poisson}) that one derives in a
dilute vortex gas. (Except for the smallest lattice,
which corresponds to a $4^3$ volume, and which can hardly
be characterised as a system at a finite $T$.)

As an aside we remark that an equally good fit can be 
obtained using the expression in (\ref{eqn_deltaS_poisson})
without the term in the denominator. Clearly much more
accurate calculations will be needed to determine
whether it is there or not. We have also tried a
fit of the form (\ref{eqn_Svortex_spinwave}) modified
by a screening factor of the form $\exp(-mL)$. This
gives a rather poor fit.

Returning to the fit using (\ref{eqn_deltaS_poisson}), we 
find that the parameter in the exponent is constrained to be
\be
a^2 \sigma(aT=0.25)  =  0.043 \pm 0.002.
\label{eqn_highT_sigfit}
\ee
We note that this is in precise agreement with the value of
the spatial string tension, $a^2 \sigma(aT=0.25) = 0.0443(4)$  
that we obtained in (\ref{eqn_sigmaT90}) 
from Polyakov loop correlations at the same value of 
$a$ and $T$. This provides a striking confirmation of
the simple vortex model. 

Can we learn something about the properties of these vortices?
For example, if we can estimate 
the vortex area and also the average density of vortices, then
we can determine whether the vortices are indeed dilute.
The density is straightforward to determine. From 
(\ref{eqn_poisson},\ref{eqn_mu_sigma}) 
the average number of vortices per unit area is
\be
{\bar n}_v  = \mu_0 = 
\frac{a^2 \sigma(T)}{2} \simeq 0.022
\ee
which implies that the typical separation between neighbouring
vortices is 
\be
\delta r = \frac{1}{\surd{\bar n}_v}  \simeq 7
\ee
in lattice units.
Now, we have previously suggested that the vortex area 
should be $A_v \sim 1/(ag^2.aT) \simeq 9$. If so, such
a vortex gas will indeed be quite dilute. 

It would be much better to estimate $A_v$ directly; but
this is not so easy because of the unknown factors, such 
as $Z_v(\beta)$. If we ignore these factors then we
can obtain wildly different estimates of $A_v$. For example,
one approach is to take the overall coefficient of the fit in 
Fig.\ref{fig_Sdiff_deconf4}
\be
2s_0 a^2 \sigma   =   0.024 \pm 0.002
\label{eqn_highT_coeffit}
\ee
and then to use (\ref{eqn_highT_sigfit}) to obtain
$s_0 \simeq 0.27$. If we use this value for $S_v$ in
(\ref{eqn_mu_poisson1}) we obtain $A_v \sim 4$. If instead
we use (\ref{eqn_mu_poisson2}) with $s_0 \simeq 0.27$
then we find $A_v \sim 70$. This huge disagreement
suggests that these calculations are much too rough 
to be useful. At a very qualitative level, the fact that
(\ref{eqn_deltaS_poisson}) fits the values in
Fig.\ref{fig_Sdiff_deconf4} all the way down to $L=6$,
with the same value of the vortex action, does suggest
that the vortex is not being squeezed in this range.
This suggests that the vortex area satisfies $A_v < 36$,
so that the vortex gas is at least moderately dilute.

\subsection{$\bf{\Delta S}$ as a function  of $\bf L$ for $\bf{T < T_c}$}
\label{subsec_dSlowT_results}

In the low $T$ confining phase one expects, from the general
analysis of
\cite{thooft}, 
that the vortex free energy, $F_v$, should approach zero
as $L \to \infty$ with a correction that 
$\propto \exp(-a^2 \sigma L^2)$. This is also the prediction of
our simple vortex model if we generalise it to a
condensate of the arbitrarily long vortex lines that would
be needed to disorder arbitrarily large Wilson loops at $T=0$.

\begin{figure}[p]
\begin{center}
\epsfig{figure=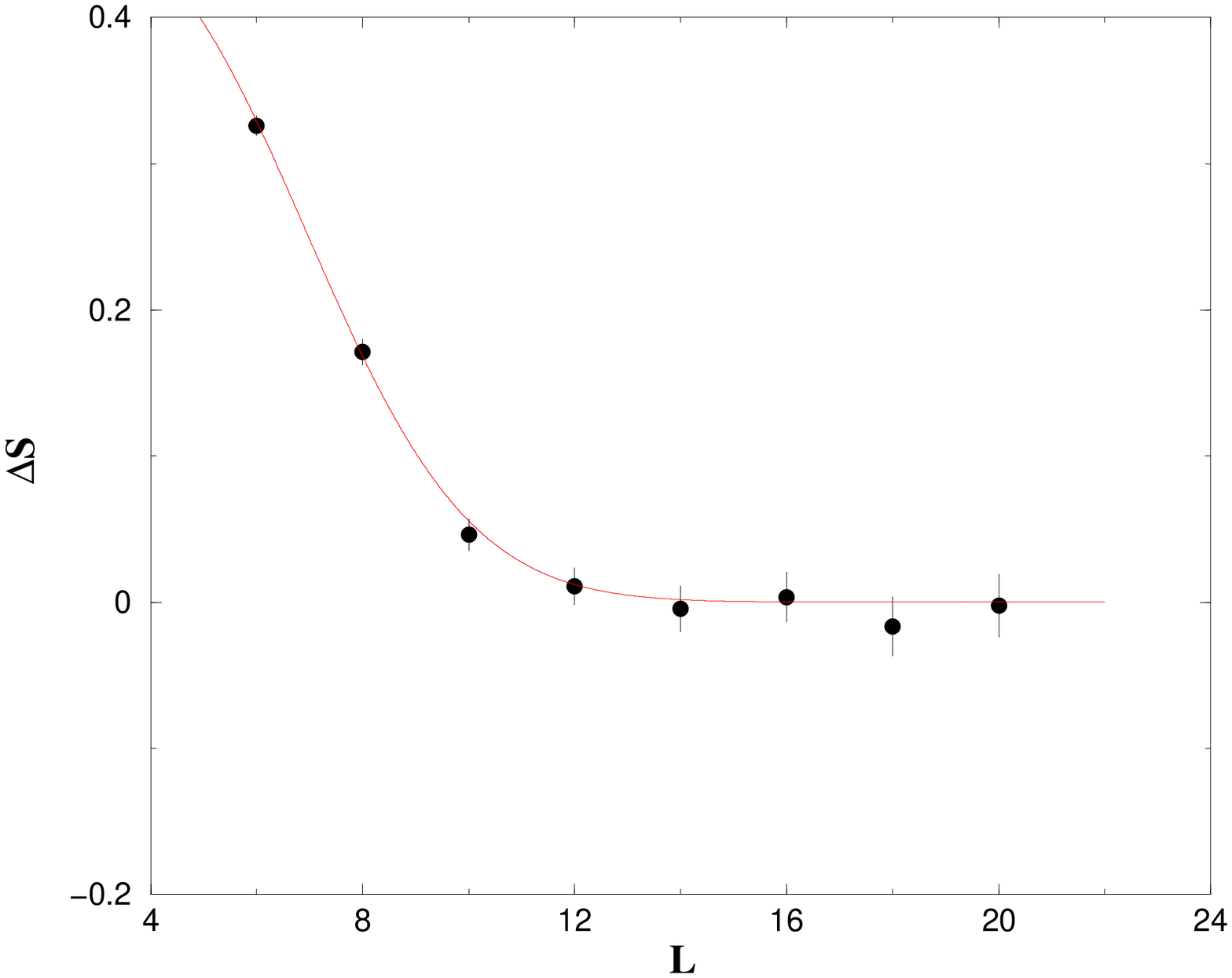, angle=270, width=15cm} 
\end{center}
\caption[]{\label{fig_Sdiff_conf6}
  {The difference of twisted and untwisted actions 
at $T=1/6a < T_c$ at $\beta = 7$, for various spatial sizes.
Fit shown is for a dilute vortex gas.
}}
\end{figure}

It would obviously be of interest to extend our high-$T$
calculations of $\Delta S$ to low $T$ so as to test these
expectations. In Fig.\ref{fig_Sdiff_conf6} we show the
result of such a calculation on lattices with $L_t=6$
at $\beta = 7$. Using (\ref{eqn_Ltcrit}) we see that this
corresponds to $T \simeq 0.7 T_c$. We also show in the
figure a best fit of the form in  (\ref{eqn_deltaS_poisson})
to the values for $L\geq 6$. This clearly fits very well. 
From the exponent we extract the value
\be
a^2 \sigma  =  \left\{ \begin{array}{ll}
0.043(2) &  \ \ \ L \geq 6 \\
0.048(4) &  \ \ \ L \geq 8
\end{array}
\right.
\label{eqn_lowT6_sigfit}
\ee
This is nicely consistent with the low-$T$ string tension,
$a^2 \sigma \simeq 0.0455$, that one obtains from (\ref{eqn_K_zeroT})
at $\beta = 7$. (The comparison with the $T=0$ string tension
is the correct one to make even though $T\not= 0$ here. The
reason is that the spatial string tension does not vary
significantly with $T$ until $T$ is close to $T_c$, as 
we see in Fig.\ref{fig_sigT_b90}.) 

As a check of the scaling properties of this behaviour,
we have also performed such a calculation on $L_t=4$
lattices at $\beta = 5$. Using (\ref{eqn_Ltcrit}) we see that 
this corresponds to $T \simeq 0.75 T_c$.  We again show the best 
fit of the form (\ref{eqn_deltaS_poisson}), in this case to 
$L\geq 4$. Again this fits very well. From the fits we extract
\be
a^2 \sigma  =  \left\{ \begin{array}{ll}
0.091(2) &  \ \ \ L \geq 4 \\
0.080(7) &  \ \ \ L \geq 6
\end{array}
\right.
\label{eqn_lowT4_sigfit}
\ee
which, we note, is reasonably consistent with the string tension,
$a^2 \sigma = 0.0979(13)$, that one obtains by a direct 
$T=0$ calculation
\cite{teper98}
at this value of $\beta$. 

So we see that  in the low $T$ confining phase, the finite 
volume corrections to the asymptotic $\Delta S = 0$ result, are 
also precisely what one expects from the simple vortex model 
(and indeed from the more general arguments of
\cite{thooft}).
We note that the fact that the functional form fits all the 
way down to $4^3$ and $6^3$ lattices at $\beta = 5$ and $7$ 
respectively, is in contrast to the high-$T$ $L_t =4$ case where 
the value of $\Delta S$ on the $4^3$ lattice falls below the 
fit. The temptation is to
interpret this as follows. As $L \downarrow$ at fixed $L_t = 1/aT$,
the lattice ceases to correspond to a thermodynamic system  at a 
fixed $T$, and so the area of the vortex begins to grow, as it 
loses its $\propto 1/T$ suppression factor. This leads to
a decrease of $s_0$ and hence of $\Delta S$. This is of course
no more than a plausible speculation.

\section{Conclusions}
\label{sec_conc}

That centre vortices might drive confinement is
an attractive idea 
\cite{thooft}
which has motivated a number of recent lattice 
calculations, e.g. 
\cite{greensite99,kovacs00}.
At high temperatures one would naively expect to find a gas of such 
vortices which, in a Euclidean calculation, would manifest themselves
as vortex world lines that close through the short ($=1/T$)
temporal boundary. In the deconfining phase these should then
provide the mechanism for `spatial' linear confinement.
The statistical mechanics of such high-$T$ vortices is relatively
simple, as we saw in a calculation that showed
how a spatial string tension, $\sigma(T)$, is generated within
a dilute gas of vortices. The observed $\sigma(T) \propto T$
behaviour arises naturally if the vortex cross-section
has an area $\propto g^2T$.  

In contrast to such expectations is the recent suggestion
\cite{cka99,cka00}
that the vortices might not be uncorrelated at large separations,
so that there is no linear spatial confinement
for $T > T_c$. To address this possibility we performed
calculations of the variation with $T$ of the vortex free energy 
and found it to be zero at all temperatures, high and low.
This supports the conventional expectation of a non-zero
spatial string tension. We then calculated the approach to
zero as the spatial volume increases and found that it
was controlled by the same spatial string tension, $\sigma(T)$, 
that had been obtained in earlier direct calculations of that
quantity. Moreover the functional behaviour is exactly like
the one we derived within the dilute vortex gas. From the
fits we estimated the mean inter-vortex distance and, much
more roughly, the vortex area. These estimates suggest that
the vortices are indeed moderately dilute. 

We performed similar calculations in the low-$T$ confining 
phase and showed that there too the finite volume
corrections are as expected for a dilute gas of
vortices, although here the argument can be made
much more generally
\cite{thooft}.

We carefully  reanalysed some earlier calculations
\cite{teper93tc,teper93kt}
of the spatial  string tension and showed that the
observed linearity of the potential extends over distances
that are large compared to the natural dynamical
length scales at high $T$.

All this provides strong evidence that the usual
expectation of an area decay for spatial Wilson loops
at high $T$ is indeed correct. It also provides
support for the notion that, at least at high $T$,
vortices do produce this area law behaviour. We
intend to pursue these questions with a better scaling 
study as well as a dedicated spatial string tension
calculation. Such studies are worth pursuing because
the dynamics of vortices should be much simpler
to identify at high $T$; both in Euclidean calculations
and  possibly through a careful
analysis of the relatively tractable D=1+1 reduced
theory.

\section*{Acknowledgments}
This work was sparked by the recent suggestion
\cite{cka99,cka00}
that the spatial string tension at high $T$ might be zero.
We are grateful to Chris Korthals Altes and Alex Kovner
for their initial encouragement as well as for 
many stimulating discussions. This study
began when two of the authors (AH and ZS) were visiting 
Oxford and they are grateful to Oxford Theoretical Physics 
for its hospitality. ZS is grateful to the Royal Society
for funding the visit, under their European Science Exchange 
Programme. His work was partially supported by the Hungarian 
National Research Fund OTKA T032501 and the Bolyai J\'anos
Research Grant. AH and BL acknowledge their postdoctoral 
funding from PPARC.

\vfill

\eject

\newpage

\begin{figure}[p]
\begin{center}
\epsfig{figure=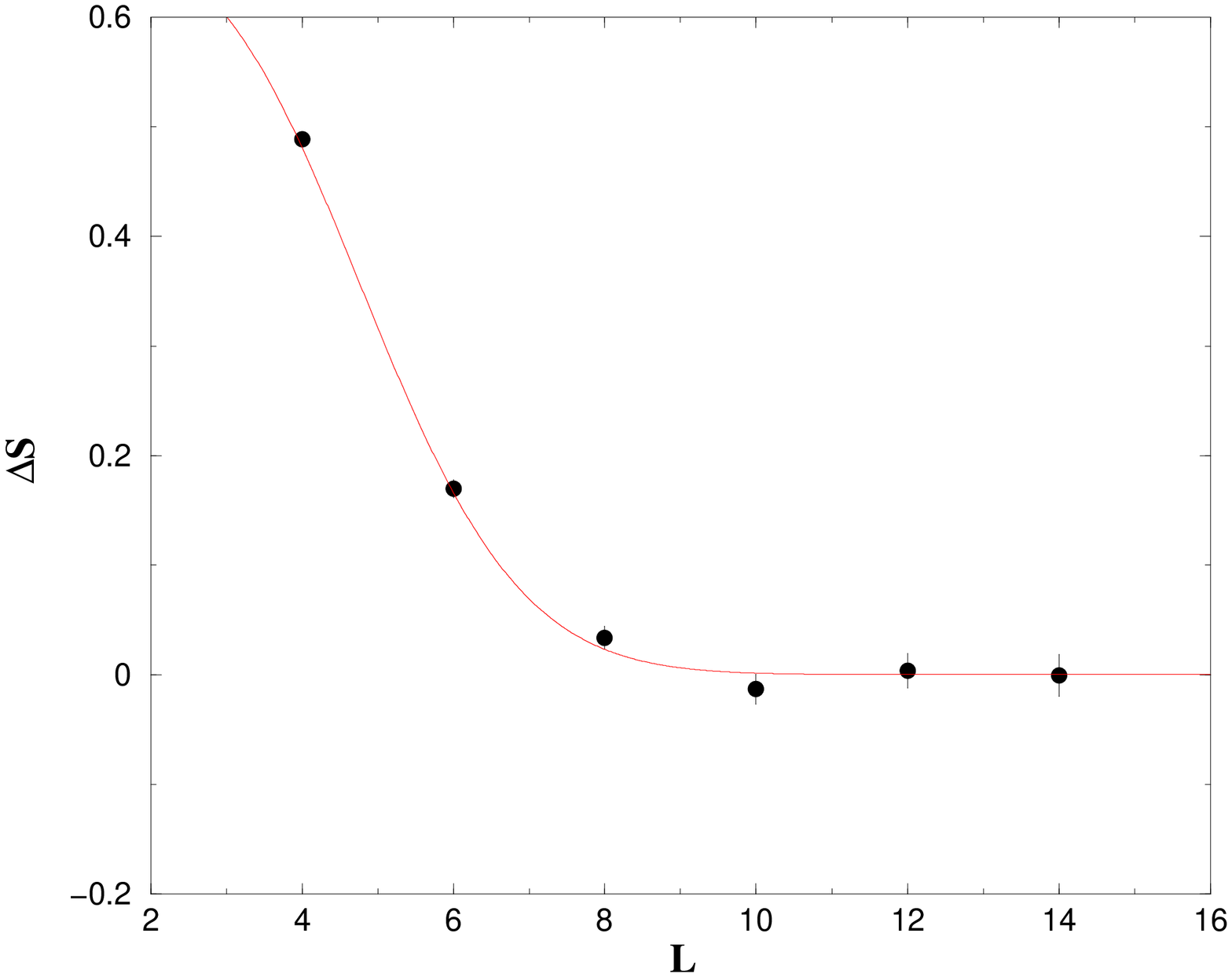, angle=270, width=15cm} 
\end{center}
\caption[]{\label{fig_Sdiff_conf4}
  {The difference of twisted and untwisted actions 
at $T=1/4a < T_c$ at $\beta = 5$, for various spatial sizes.
Fit shown is for a dilute vortex gas.
}}
\end{figure}

\end{document}